\documentclass[]{emulateapj}
\usepackage{graphicx}
\usepackage{hhline}
\usepackage{xcolor}
\usepackage{amssymb, amsmath}

\newcommand{\Ms}{M_{\odot}}

\newcommand{\leff}{\widetilde\Lambda}
\newcommand{\mc}{\mathcal{M}_c}

\begin{document}

\title{Tidal deformability from GW170817 as a direct probe of the neutron star radius}
\author{Carolyn A. Raithel, Feryal \"Ozel, \& Dimitrios Psaltis}
\affiliation{Department of Astronomy and Steward Observatory, University of Arizona, 933 N. Cherry Avenue, Tucson, Arizona 85721, USA}

\begin{abstract}
Gravitational waves from the coalescence of two neutron stars were recently detected for the first time by the LIGO-Virgo collaboration, in event GW170817. This detection placed an upper limit on the effective tidal deformability of the two neutron stars and tightly constrained the chirp mass of the system. We report here on a new simplification that arises in the effective tidal deformability of the binary, when the chirp mass is specified. We find that, in this case, the effective tidal deformability of the binary is surprisingly independent of the component masses of the individual neutron stars, and instead depends primarily on the ratio of the chirp mass to the neutron star radius. Thus, a measurement of the effective tidal deformability can be used to directly measure the neutron star radius. We find that the upper limit on the effective tidal deformability from GW170817 implies that the radius cannot be larger than $\sim$13~km, at the 90\% level, independent of the assumed masses for the component stars. The result can be applied generally, to probe the stellar radii in any neutron star-neutron star merger with a measured chirp mass. The approximate mass-independence disappears for neutron star-black hole mergers. Finally, we discuss a Bayesian inference of the equation of state that uses the measured chirp mass and tidal deformability from GW170817 combined with nuclear and astrophysical priors and discuss possible statistical biases in this inference. 
\end{abstract}

\maketitle

\section{Introduction}

The first detection of gravitational waves from a neutron star-neutron star merger (GW170817, \citealt{Abbott2017a}) marks the start of a new era in the study of neutron stars, their associated transient events, and the dense-matter equation of state. The electromagnetic counterpart that accompanied the event \citep{Abbott2017b} has confirmed neutron star mergers as the sources of at least some short-duration gamma-ray bursts, as has long been theorized \citep{Eichler1989, Narayan1992, Berger2014}, as well as the source of kilonovae, predicted to be powered by the radioactive decay of merger ejecta \citep{Li1998, Metzger2010}. Information about the component neutron stars and their underlying equation of state is encoded in the waveform itself, which was observed by the two LIGO and one Virgo detectors for $\sim$3000 orbital cycles prior to the merger \citep{Abbott2017a}.

Several studies have already placed constraints on fundamental neutron star properties using these observations. For example, \citet{Margalit2017} used the combined gravitational wave and electromagnetic signals to set an upper limit on the maximum neutron star mass, which is a sensitive constraint on the equation of state at high densities \citep{Ozel2009}. In another work,  \citet{Rezzolla2017} inferred the maximum neutron star mass from the event without relying on models of the electromagnetic signal, instead using only the quasi-universal relations that describe neutron stars and simple models of kilonovae.

The observed gravitational waveform can also be used to place direct constraints on the neutron star equation of state (EOS). In one of the first quantitative studies exploring EOS effects on the waveform from the coalescence of two neutron stars, \citet{Read2009a} showed that a realistic waveform would deviate significantly from a point-particle waveform and that this could be observed with Advanced LIGO. The degree of the deviation depends on the underlying EOS and, as a result, could be used to differentiate between EOS that differ in radius by only $\sim$1~km \citep{Read2009a, Read2013, Lackey2015}.

The magnitude of the deviation is strongest at later times in the inspiral and during the merger, i.e., in the phases where numerical relativity would be necessary to model the waveforms. Nevertheless,  \citet{Flanagan2008} found that the early phase of the inspiral depends cleanly on a single EOS-dependent parameter: the tidal Love number, $\lambda$. The tidal Love number measures the ratio of the star's tidally-induced quadrupolar deformation, $Q^{\rm(tid)}$, to the tidal potential caused by a binary companion, $\varepsilon^{\rm(tid)}$, i.e.,
\begin{equation}
\label{eq:love}
\lambda \equiv - \frac{Q^{(\rm tid)}}{\varepsilon^{(\rm tid)}}   
\end{equation}
or, in dimensionless form,
\begin{equation}
\label{eq:loveDimless}
\Lambda \equiv \frac{\lambda}{M^5} \equiv \frac{2}{3} k_2^{(\rm tid)}\left( \frac{R c^2}{G M} \right)^5,
\end{equation}
where $R$ is the radius of the neutron star and $M$ is its mass. Following the convention of \citet{Flanagan2008}, we call $k_2$ the tidal apsidal constant. The tidal apsidal constant depends both on the equation of state and the compactness ($GM/Rc^2)$ of the particular star. For realistic, hadronic equations of state, $k_2$ has been constrained to lie in the range $\sim 0.05-0.15$ \citep{Hinderer2008, Hinderer2010, Postnikov2010}.

The individual Love numbers for the two stars, $\Lambda_1$ and $\Lambda_2$, cannot be disentangled in the observed gravitational waveform. Instead, what is measured is an effective tidal deformability of the binary, $\leff$, which is a mass-weighted average of $\Lambda_1$ and $\Lambda_2$ that we describe in detail in $\S$\ref{sec:properties}. The expectation is thus that $\leff$ would measure a mass-weighted compactness for the two neutron stars. Similarly, the two component masses are not measured directly; rather, the chirp mass is.

We report here on a new simplification that arises in the effective tidal deformability of the binary when the chirp mass is measured accurately. We find that $\leff$ depends primarily on the ratio of the chirp mass to the neutron star radius. Thus, we find that $\leff$ can be used as a direct probe of the neutron star radius, rather than of the compactness as is typically assumed.

In $\S$\ref{sec:properties}, we describe the measured properties of GW170817. We show in $\S$\ref{sec:leff} that the effective tidal deformability is approximately independent of the component masses, when the chirp mass is specified. In $\S$\ref{sec:newtonian}, we use the Newtonian limit to show analytically that the mass-independence arises from an inherent symmetry in the expression for the effective tidal deformability. Finally, in $\S$\ref{sec:bayes}, we perform an example Bayesian inference of the neutron star EOS from the measured tidal deformability and chirp mass and a limited number of prior physical constraints and discuss important statistical biases that can occur in such inference schemes.

\section{Properties of GW170817}
\label{sec:properties}
The properties of GW170817 were inferred by matching the observed waveform with a frequency-domain post-Newtonian waveform model \citep{Sathyaprakash1991}, with modifications to account for tidal interactions \citep{Vines2011}, point-mass spin-spin interactions \citep{Mikoczi2005, Arun2011, Bohe2015, Mishra2016}, and effects due to spin-orbit coupling \citep{Bohe2013}. The LIGO analysis using these models is summarized in \citet{Abbott2017a} and references therein.

One of the most tightly constrained properties that was inferred is the chirp mass, defined as
\begin{equation}
\mc = \frac{(m_1 m_2)^{3/5} }{(m_1 + m_2)^{1/5}} = m_1 \frac{q^{3/5}}{(1+q)^{1/5}},
\label{eq:mc}
\end{equation} 
where $m_1$ and $m_2$ are the masses of the primary and the secondary neutron stars, respectively, and we have introduced the mass ratio, $q \equiv m_2/m_1$. The chirp mass was constrained to $\mc = 1.188\substack{+0.004 \\ -0.002}~\Ms$ at the 90\% confidence level, independent of the particular waveform model or priors chosen \citep{Abbott2017a}.

By assuming low-spin priors, as is consistent with the binary neutron star systems that have been observed in our Galaxy, the component masses were inferred from the chirp mass to lie within the ranges  $m_1 \in (1.36, 1.60)~\Ms$ and $m_2 \in (1.17, 1.36)~\Ms$, with a mass ratio of $q \in (0.7, 1.0)$, all at the 90\% confidence level \citep{Abbott2017a}. These masses are consistent with the range of masses observed masses in other neutron star systems (see \citealt{Ozel2016} for a recent review of neutron star mass measurements). 

GW170817 also provided constraints on the effective tidal deformability of the system, defined as
\begin{equation}
\leff \equiv \frac{16}{13} \frac{ (m_1 +  12 m_2) m_1^{4} \Lambda_1 + (m_2 + 12 m_1) m_2^{4} \Lambda_2}{(m_1+m_2)^5},
\label{eq:leff}
\end{equation}
\citep{Flanagan2008, Favata2014}. In equation~(\ref{eq:loveDimless}), we saw that the dimensionless tidal Love number depends only on the stellar compactness and the tidal apsidal constant, which in turn depends on the equation of state and compactness. Combining these expressions, we can explicitly write the dependence of the effective tidal deformability on neutron star properties as $\leff = \leff(m_1, m_2, R_1, R_2, \rm EOS)$. 

\citet{Abbott2017a} constrain the effective tidal deformability for GW170817 to be $\leff \le 800$ at the 90\% confidence level, which disfavors EOS that predict the largest radii stars. In the following analysis, we will show that this measurement can also be used to directly constrain the radii of the individual neutron stars, independently of the component masses.

\section{Effective tidal deformability for GW170817}
\label{sec:leff}
\begin{figure}[ht]
\centering
\includegraphics[width=0.45\textwidth]{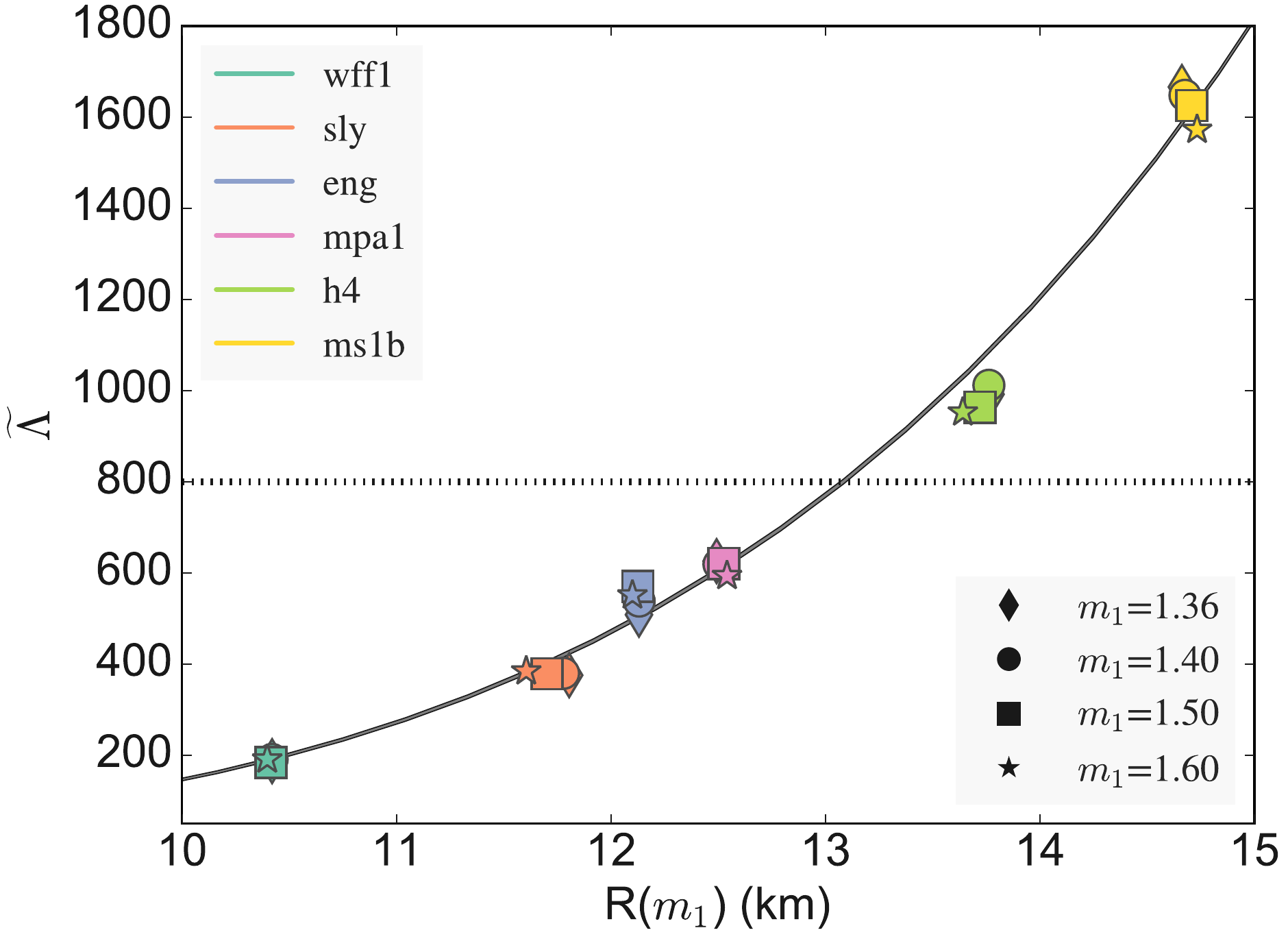}
\caption{\label{fig:LvsR} Effective tidal deformability of the binary system as a function of the radius of the primary neutron star. The tidal deformability is calculated for various primary masses (corresponding to the different symbols) using several proposed equations of state (corresponding to the different colors). The mass of the secondary neutron star is found assuming the chirp mass, $\mc$ = 1.188~$\Ms$, from GW170817. The observed 90\%-confidence upper limit on $\leff\le 800$ is shown as the dotted line. The narrow band (which is indistinguishable from a single curve) shows the range for $q=0.7-1.0$ from equation~(\ref{eq:expansion}). We find that $\leff$ is relatively insensitive to $m_1$ but scales strongly with radius, and that the upper limit for GW170817 implies $R\lesssim13$~km.}
\end{figure}

We start with a simple illustration of our key result. Figure~\ref{fig:LvsR} shows the effective tidal deformabilities as a function of the stellar radii for a number of realistic EOS. For each EOS, we calculated these tidal deformabilities for various values of $m_1$ that lie within the mass range inferred for GW170817 (shown in different symbols). The corresponding values for $m_2$ are calculated assuming a fixed chirp mass, $\mc=1.188~\Ms$.  

We find that $\leff$ is almost entirely insensitive to the mass of the component stars for the relevant mass range and depends instead primarily on the radius of the star. In particular, $\leff$ changes by nearly an order of magnitude between $R=10$~km and $R=15$~km, but, for a given radius, changes negligibly for masses spanning the full range of $m_1=1.36-1.6~\Ms$. 

An upper limit of $\leff \lesssim 800$ immediately excludes radii above $\sim$13~km at the 90\% confidence level, without requiring detailed knowledge of $m_1$. As shown in Figure~\ref{fig:LvsR}, this rules out the EOS that predict the largest radii, such as the hyperonic EOS H4 \citep{Lackey2006} and the field theoretic nucleonic EOS with a low symmetry energy of 25 MeV, MS1b \citep{Muller1996}.

The trend found in Figure~\ref{fig:LvsR} is for a sample of six EOS. However, this result is more general, as we will now show. It has been reported previously that the individual tidal deformabilities of neutron stars obey a universal relationship with stellar compactness \citep{Yagi2013}. In particular,  \citet{Yagi2017} found that the relationship can be written as
\begin{equation}
\label{eq:YY}
C = a_0 + a_1 \ln\Lambda + a_2 (\ln\Lambda)^2,
\end{equation}
where $C \equiv GM/Rc^2$ is the compactness and the coefficients were fit to be $a_0=0.360, a_1=-0.0355$, and $a_2=0.000705$. The relation holds to within 6.5\% for a wide variety of neutron star EOS \citep{Yagi2017}. 

\begin{figure}[ht]
\centering
\includegraphics[width=0.42\textwidth]{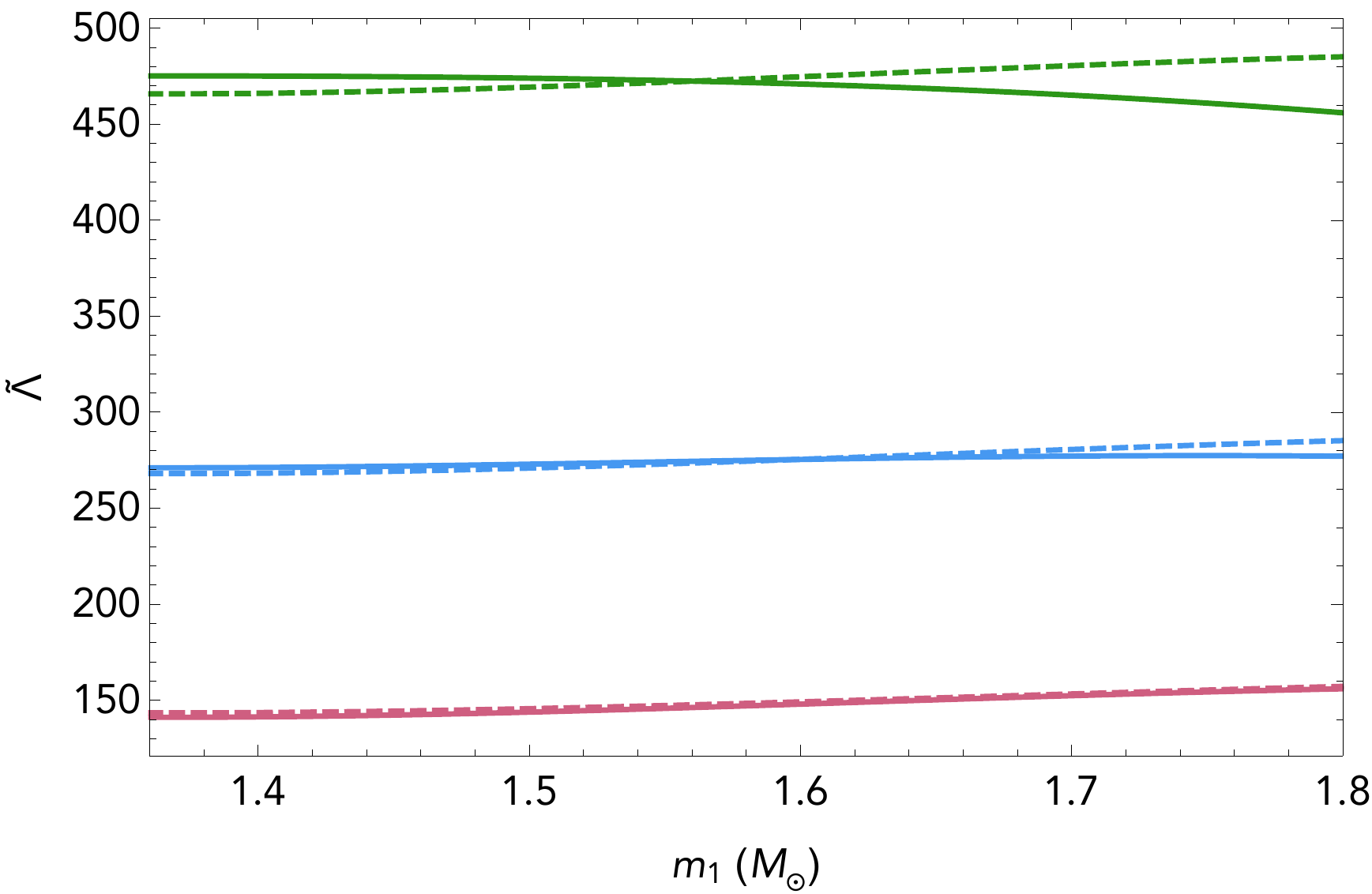}
\caption{\label{fig:LvsM} Effective tidal deformability of the binary system as a function of the primary mass, $m_1$, when the chirp mass is held fixed at $\mc=1.188~\Ms$. We calculate $\leff$ for three fixed radii, R=10, 11, and 12~km, shown in purple, blue, and green, respectively. The solid lines show the tidal deformability calculated using the empirically-fit universal relation between the tidal deformability of each neutron star and its compactness from \citet{Yagi2017}, while the dashed lines show the quasi-Newtonian approximation for  $\Lambda_i$ from equation~(\ref{eq:lambda}). The quasi-Newtonian approximation is a good approximation to the full GR result. }
\end{figure}

To see if the trend we have found between $\leff$ and $R$ holds generically for a wide range of EOS, we use the universal relation of equation~(\ref{eq:YY}) to calculate the individual tidal deformabilities, $\Lambda_1$ and $\Lambda_2$.  We then calculate the effective tidal deformability for the binary system, shown as the solid lines in Figure~\ref{fig:LvsM} for three different radii. We find that when we use this universal relation to represent a much larger sample of EOS, the trend holds. The effective tidal deformability of the binary depends extremely weakly on the component masses but strongly on the radii of the stars.

The weak dependence of $\leff$ on the component masses is surprising and has not been reported before. It renders $\leff$ a direct probe of the neutron star radius, rather than of the compactness as is typically assumed. We turn now to an analytic explanation of the origin of this result.

\section{Effective tidal deformability in the Newtonian limit}
\label{sec:newtonian}

In order to see why the dependence on mass in equation~(\ref{eq:leff}) for $\leff$ is so weak, we turn to the Newtonian limit. \citet{Yagi2013} showed that the Newtonian expression for the tidal Love number of a star governed by a polytropic EOS with index $n=1$ (which is appropriate for the majority of realistic EOS) is simply
\begin{equation}
\Lambda_N = \frac{15-\pi^2}{3 \pi^2} \left(\frac{Rc^2}{Gm}\right)^5.
\end{equation}

The full relativistic expression for the tidal deformability of a star is given by \citet{Damour2009} for a given compactness and a parameter $y$, which is the logarithmic derivative of a metric function, $H$, at the stellar surface. The full expression is far more complicated than what we have introduced so far, but we find that a relatively simple metric correction to $\Lambda_N$ qualitatively reproduces the universal results computed for more realistic EOS. We call this correction the ``quasi-Newtonian" expression and define it as
\begin{equation}
\label{eq:lambda}
\Lambda_{qN } = \frac{15-\pi^2}{3 \pi^2} \left(\frac{R c^2}{Gm} \sqrt{1-\frac{2 G m}{R c^2}} \right)^5.
\end{equation}
This corresponds to equation (96) of \citet{Damour2009} with $\beta \approx 1$.

We can combine this with equation~(\ref{eq:leff}) to write the quasi-Newtonian effective tidal deformability as
\begin{multline}
\label{eq:leffN}
\leff_{qN} = \frac{16}{13} \frac{15-\pi^2}{3 \pi^2}\left( \frac{R c^2}{G m_1}\right)^5 \times \\
 \frac{ (1 +  12 q)\left(1-\frac{2 G m_1}{R c^2}\right)^{5/2} + (1 + 12/q)\left(1-\frac{2 G q m_1}{R c^2}\right)^{5/2} }{(1+q)^5} ,
\end{multline}
where we have assumed that the radii for the two neutron stars are the same, as is approximately true for $n=1$ polytropic EOS. Finally, we can eliminate $m_1$ in favor of $\mc$ and $q$ using equation~(\ref{eq:mc}), yielding an expression for $\leff_{qN}$ in terms of only $q$, $\mc$, and $R$.

This quasi-Newtonian form of $\leff_{qN}$  is much simpler to work with, but is it a good enough approximation? We show $\leff$ and $\leff_{qN}$ as functions of $m_1$ in Figure~\ref{fig:LvsM} as the solid and dashed lines, respectively, for fixed radii of R=10, 11, and 12~km and fixed $\mc=1.188~\Ms$.  We find that the quasi-Newtonian approximation provides a reasonable approximation of the full expression for $\leff$, calculated using the quasi-universal relations. We can, therefore, use $\leff_{qN}$ to understand its dependence on the masses.

Expressing $\leff_{qN}$ as a series expansion around $q=1$, we find
\begin{equation}
\label{eq:expansion}
\leff_{qN} = \leff_{0} \left( 1 + \delta_{0} (1-q)^2\right) + \mathcal{O}\left((1-q)^3\right),
\end{equation}
where
\begin{equation}
\label{eq:coef}
\leff_{0}  = \frac{15-\pi^2}{3 \pi^2} \xi^{-5} (1-2 \xi)^{5/2},
\end{equation}
\begin{equation}
\label{eq:correction}
\delta_{0} =  \frac{3}{104}(1-2 \xi )^{-2}\left(-10 + 94 \xi  - 83 \xi^2 \right),
\end{equation}
and we have introduced 
\begin{equation}
\label{eq:xi}
\xi = \frac{2^{1/5} G \mc}{ R c^2}
\end{equation}
as an ``effective compactness."

We note that expanding near $q=1$ is not a restrictive choice. The known population of neutron stars is observed to have a relatively small range of masses and the observed mass distribution of double neutron stars is even narrower, suggesting that most astrophysical merger scenarios will have $q$ near unity (see \citealt{Ozel2016}).

From eqs.~(\ref{eq:expansion}-\ref{eq:xi}), we see that the effective tidal deformability of the binary, $\leff$, scales approximately as $R^5$ for a given $\mc$. When the mass ratio is close to unity, the individual masses add only a small correction. For the measured chirp mass of GW170817, we calculate the expansion coefficients for a few radii in Table~\ref{table:expansion}. We note that the mass dependence only enters at order $(1-q)^2$. Furthermore, the weak dependence on mass becomes even weaker as the radius increases. Even for $R=10$~km, the mass dependent term adds at most a $\sim$4\% correction to $\leff_{qN}$ for the mass ratio range inferred for GW170817.

We show this quasi-Newtonian expansion for a range of q values, $q \in (0.7,1.0)$, as the narrow gray band in Figure~\ref{fig:LvsR} and find that it does recreate the trend observed in that sample of EOS.

\begin{deluxetable}{ccc}[ht]
\tabletypesize{\footnotesize}
\tablewidth{0.45\textwidth}
\tablecaption{$\leff_{qN}$ expansion terms for the chirp mass measured from GW170817. }
\tablehead{
\colhead{Radius} & 
\colhead{$\leff_0$} &
\colhead{Expansion } 
} 
\startdata
$R=10$~km & 143.4 & $1 + 0.041 \left(\frac{1-q}{1-0.7}\right)^2 + \mathcal{O}\left( \frac{1-q}{1-0.7}\right)^3$ \\
$R=11$~km & 268.0 & $1 + 0.029 \left(\frac{1-q}{1-0.7}\right)^2 + \mathcal{O}\left( \frac{1-q}{1-0.7}\right)^3$ \\
$R=12$~km & 465.8 & $1 + 0.020 \left(\frac{1-q}{1-0.7}\right)^2 + \mathcal{O}\left( \frac{1-q}{1-0.7}\right)^3$ \\
$R=13$~km & 764.6 & $1 + 0.014  \left(\frac{1-q}{1-0.7}\right)^2 + \mathcal{O}\left( \frac{1-q}{1-0.7}\right)^3$ \\
\enddata
  \label{table:expansion}
\end{deluxetable}

\subsection{Black hole-neutron star mergers}
\label{sec:bh}
Black hole-neutron star mergers are another source of gravitational waves that may contain information about the neutron star EOS. The tidal Love number of a black hole is zero \citep{Damour2009, Binnington2009}, which greatly simplifies the effective tidal deformability of equation~(\ref{eq:leff}). However, this simplification also destroys the inherent symmetry in equation~(\ref{eq:leff}), which is the source of the mass independence in the neutron star-neutron star merger scenario. Without this symmetry, a series expansion of $\leff$, as in equation~(\ref{eq:expansion}), includes a correction term of order $(1-q)$.

Due to the lower-order terms of $\mathcal{O}(1-q)$, there persists a stronger dependence on the mass of the components. Thus, the effective tidal deformability measured from a neutron star-black hole merger does not directly probe the radius, as in the case of a neutron star-neutron star merger. Instead, a measurement of $\leff$ will primarily probe the neutron-star compactness.

\section{Bayesian inference of the radius}
\label{sec:bayes}

\begin{figure*}[ht]
\centering
\includegraphics[width=0.95 \textwidth]{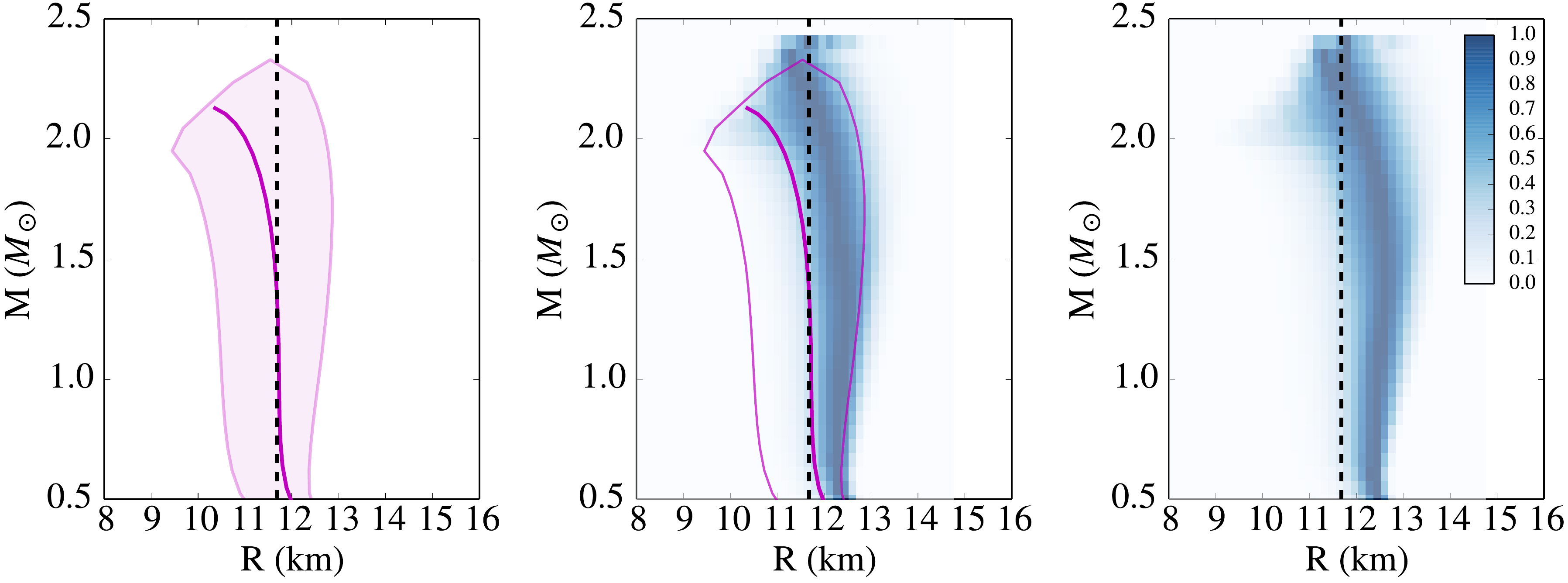}
\caption{\label{fig:EOS_MR} Left: Mass-radius relations corresponding to the most-likely EOS in our Bayesian inference, with a sample distribution for $\leff$ centered at $\leff=400$ and a fixed chirp mass of $\mc=1.188~\Ms$. The solid magenta line corresponds to the most-likely EOS, while the pink band corresponds to the range of EOS with posteriors within $1/\sqrt{e}$ of the maximum value. The black dashed line shows the analytic prediction from our $\leff-R$ relation of equation~(\ref{eq:expansion}). We find excellent agreement between our $\leff-R$ prediction and the full Bayesian inference. Middle: Same as left panel, but showing, in addition, the marginalized posteriors over the neutron star radii for a fixed grid of masses. These marginalized likelihoods are shown in blue.  By marginalizing the posteriors in this way, the results are skewed to higher radii and away from the maximum likelihood solution. 
Right: Marginalized likelihoods for an inference with only the priors and no data. These marginalized posteriors are nearly identical to the marginalized posteriors from the inference that incorporated data from a $\leff=400$ centered Gaussian. This method of marginalization over-weights the prior on the EOS pressures imposed by the observation of a 1.97~$\Ms$ neutron star. The results of the marginalization are less sensitive to the input data and are not reliable.}
\end{figure*}

In Figure~\ref{fig:LvsR}, we showed that $\leff$ can be used to directly probe the neutron star radius. The measurement from GW170817 of $\leff \le$~800, at the 90\% confidence level, already implies that the radii of the neutron stars should be $\lesssim$~13~km. However, in order to place more robust constraints on the neutron star radius or to place comprehensive constraints on the underlying EOS, we need to incorporate prior physical constraints and other observations within a full Bayesian framework.

\citet{Read2009} and \citet{Ozel2009} introduced the use of piecewise polytropic EOS to convert the observations of neutron stars into constraints on the EOS in a statistically robust way. In \citet{Raithel2016}, we showed that an optimal parametrization of the neutron star EOS, given the expected accuracy of measurements in the near future, requires 5 piecewise polytropes. In \citet{Raithel2017}, we further developed the statistical framework with which to perform a Bayesian inference of the pressures of our parametric EOS. In our inference here, we incorporate a variety of astrophysical and nuclear physics priors, including that the EOS is microscopically stable and causal at all pressures, that the lowest two pressures exceed the limit placed by two-nucleon interaction, and that all EOS must produce a neutron star of at least 1.97~$\Ms$, in order to be within $1\sigma$ of the measurements of the most massive neutron stars \citep{Antoniadis2013, Fonseca2016}. We assume a uniform prior on the pressures. In order not to over-parametrize the EOS, while still allowing the possibility of complex behavior to be inferred, we also include a Gaussian regularizer over the second derivative of the pressure ($\lambda=2$), which penalizes sharp phase transitions. For further details on the set-up of our Bayesian inference, see \citet{Raithel2017}. 

In addition to the above priors, which were extensively studied in \citet{Raithel2017}, we also place an \textit{upper} limit on the maximum mass, $M_{\rm max} < 2.33~\Ms$, which is the upper limit of the 90\% credibility level found in \citet{Rezzolla2017}. This maximum mass was inferred from GW170817 assuming only the quasi-universal neutron star relations and simple models of kilonovae and is thus fairly model-independent.

Our goal is to perform a sample Bayesian inference, using the type of data that came from GW170817. Unfortunately, because only an upper limit on $\leff$ was provided by the LIGO-Virgo Collaboration, rather than the full posterior information, we can only perform example inferences at this point.  

For the sample inference here, we take the constraint on $\leff \le 800$ to correspond to a Gaussian distribution, centered at $\leff_{\rm obs}=400$ with a dispersion of $\sigma_{\leff}=243$. We also use the inferred chirp mass from GW170817, which is constrained to $\mc = 1.188\substack{+0.004 \\ -0.002}~\Ms$.

The likelihood of a particular EOS is given by
\begin{equation}
\label{eq:prob_full}
P(\mathrm{ EOS | \mc, \leff}) = P_{pr}( \mathrm{EOS}) P(\mc, \leff | \mathrm{ EOS} ),
\end{equation}
where $P_{pr}(\rm EOS)$ represents the set of the priors on the EOS, which we describe above. Because of the high accuracy in the measurement of the chirp mass, we fix it to the observed value, and use that to set $m_2$ for any given $m_1$. Then, equation~(\ref{eq:prob_full}) can be written as

\begin{multline}
\label{eq:prob}
P(\mathrm{ EOS | \leff}) = P_{pr}( \mathrm{EOS}) \times \\
  \frac{1}{\sqrt{2 \pi} \sigma_{\leff}}  \exp\Big\{ -\frac{[\leff_{\rm EOS(m_1, m_2)}-\leff_{\rm obs}]^2}{2 \sigma_{\leff}^2} \Big\},
\end{multline}
where $\leff_{\rm EOS(m_1, m_2)}$ is the effective tidal deformability for a particular set of the two masses,  $m_1$ and $m_2$, of each EOS that maximizes the likelihood. We choose to use the maximum likelihood, rather than integrating over all combinations of $m_1$ and $m_2$ to avoid biasing our results, as discussed in \citet{Raithel2017}.

To populate the posteriors in equation~\ref{eq:prob}, we run a Markov Chain Monte Carlo (MCMC) simulation with $\sim10^6$ points. For each EOS that is tested in our MCMC, we also calculate the corresponding mass-radius relation using the standard TOV equations. In the left panel Figure~\ref{fig:EOS_MR}, we show the mass-radius relations corresponding to the highest-likelihood solutions from our MCMC. The solid magenta line shows the most likely solution, while the pink shaded band corresponds to the range of EOS with probabilities within $1/\sqrt{e}$ of the maximum value.  Figure~\ref{fig:EOS_MR} also shows, as the black dashed line, the radius that corresponds to the most likely value of $\leff=400$ using the quasi-universal relation of equation~(\ref{eq:expansion}). Both our analytic expansion of the $\leff-$radius relationship and the full Bayesian inference presented here imply radii of $\sim11.7$~km, for these sample data. This Bayesian method can be used to robustly infer the EOS as additional measurements of $\leff$ and $\mc$ are made from future neutron star merger events.

As a note of caution, we show in the middle and right panels of Figure~\ref{fig:EOS_MR} the results of our MCMC after they have been marginalized in mass-radius space, as is frequently presented in some other studies \citep[e.g.,][]{Steiner2017, Most2018}. This method of marginalization involves calculating the posteriors over radius in a fixed grid of masses. However, because there are far more large-radii EOS that produce a 2~$\Ms$ neutron star, marginalizing in this way effectively weights the large radii solutions much more heavily than any other priors, or even than the data themselves. This can be seen in the middle panel of Figure~\ref{fig:EOS_MR}, which shows that the marginalized solution leads to an inferred radius of $\sim$12.2~km, even though the maximum likelihood solution occurs at $\sim$11.7~km. To further illustrate the point, we show in the right panel of Figure~\ref{fig:EOS_MR} the marginalized posteriors for an inference with only priors and no data at all. The marginalized posteriors with no data are effectively identical to the marginalized posteriors for the inference that incorporated data from a $\leff=400$ centered Gaussian. This method of marginalization weights the 2~$\Ms$ prior so heavily that the data are effectively ignored. We suspect that this bias also affects the posteriors presented in other works, e.g. \citet{Most2018}. For further discussion of the bias introduced by such a marginalization, see \citet{Raithel2017}.

\section{Conclusions}

In this paper, we found that the effective tidal deformability is approximately independent of the component masses for a neutron star-neutron star merger, when the chirp mass is specified. Because this surprising result is difficult to see analytically in the full GR case, we introduce a quasi-Newtonian approximation that closely reproduces the results found in full GR. In the quasi-Newtonian limit, we find that the masses of the stars only enter at order $\mathcal{O}\left((1-q)^2\right)$, where $q$ is the mass ratio. We find that, for the chirp mass measured from GW170817, this introduces at most a 4\% mass correction to the effective tidal deformability for the entire range of mass ratios. Thus, the effective tidal deformability can be considered as approximately independent of the neutron star masses. This makes $\leff$ a direct probe of the neutron star radius. For GW170817, we find that the 90\% upper limit on $\leff$ implies that the neutron star radius must be $\lesssim$13~km. 

In the case of a neutron star-black hole merger, we find that the vanishing $\Lambda$ for the black hole breaks the symmetry in $\leff$ and makes it depend more strongly on the component masses. Thus, a measurement of $\leff$ for a neutron star-black hole merger probes the compactness of the neutron star, but cannot be used as a direct probe of the radius.

Finally, we incorporate other astrophysical priors and constraints from nuclear physics in order to perform an example Bayesian inference of the pressures in a parametric EOS, from the $\mc$ value inferred in GW170817 and a sample interpretation of the reported upper limit on $\leff$. We find that, even when such priors are included, we infer a mass-radius relation that is consistent with the analytic prediction from our $\leff-R$ universal relationship.  We show that significant biases can be avoided by robustly examining the maximum likelihood solutions in the multi-dimensional parameter space, rather than introducing a marginalization in mass-radius space. The marginalization tends to weight particular priors more heavily than the actual data, which causes the resulting answer to skew systematically towards larger radii.

Using the methods we have developed in this paper, future gravitational  wave events can be used to directly and robustly constrain the neutron star radius, providing new constraints on the EOS.

{\em{Acknowledgements.\/}} We thank Sam Gralla for useful discussions on this work. CR is supported by the NSF Graduate Research Fellowship Program Grant DGE-1143953. FO and DP are supported by NASA grant NNX16AC56G.

\bibliography{gwbib}
\bibliographystyle{apj}

\end{document}